\documentclass{article}

\usepackage[english]{babel}

\usepackage[letterpaper,top=2cm,bottom=2cm,left=3cm,right=3cm,marginparwidth=1.75cm]{geometry}
\usepackage{xcolor}

\usepackage{amsmath}
\usepackage{graphicx}
\usepackage{url}
\usepackage{breakurl}
\usepackage[colorlinks=true, linkcolor=blue, urlcolor=teal, citecolor=cyan,breaklinks=true]{hyperref}
\graphicspath{ {./images/} }

\title{
{\bf Enabling the Digital Democratic Revival}: \\
A Research Program for Digital Democracy
}

\author{Lorentz Center Workshop on ``\href{https://www.lorentzcenter.nl/algorithmic-technology-for-democracy.html}{Algorithmic Technology for Democracy}'' \\
The list of authors follows at the end of the paper
}

\begin{document}
\maketitle

\begin{abstract}
This white paper outlines a long-term scientific vision for the development of digital-democracy technology. We contend that if digital democracy is to meet the  ambition of enabling a participatory renewal in our societies, then a comprehensive multi-methods research effort is required that could, over the years, support its development in a democratically principled, empirically and computationally informed way. The paper is co-authored by an international and interdisciplinary team of researchers and arose from the Lorentz Center Workshop on \href{https://www.lorentzcenter.nl/algorithmic-technology-for-democracy.html}{``Algorithmic Technology for Democracy''} (Leiden, October 2022). 
\smallskip

For comments, suggestions, improvements, criticisms please reach out to the coordinating authors listed at the end of the paper. We plan to update the document regularly.
\end{abstract}


\tableofcontents

\section{Digital tools for a participatory democratic revival?
}
\paragraph{Democratic crisis and participatory democratic revival}
Democracy is under pressure. 
Citizens' dissatisfaction with democracy has been running high for decades \cite{foa2020global}, and 
one of its causes 
is the perception of a growing disconnect between political decision makers and citizens \cite{gilens2005inequality,fiorina2012disconnect,wagner2022yellow,pew2}. Many citizens find themselves distrusting democratic institutions, feeling sidelined rather than represented by political elites perceived to pursue the needs of various interest groups over those of society. A chief route to counter these feelings and strengthen democracy is to better include citizens throughout the deliberation and decision-making processes behind policy decisions \cite{matsusaka2020let,pena2020innovative}.

However, most existing approaches to civic engagement either include tiny numbers of citizens (as in citizen assemblies) or exclude active deliberation and limit participation to the support of petitions (as in e-petitions). Furthermore, they typically fail to explicate a clear and transparent path from citizens' input to political action that is required for any democratic process \cite{fishkin2021deliberation}. 
In this position paper, we echo earlier calls for the deployment of digital tools to overcome the limitations of existing approaches to civic engagement. Digital tools are uniquely suited to address these shortcomings of current approaches, as they allow for the development of scalable and interactive methods deployable at various levels of governance: from local, to national and even global. However, we also identify remaining roadblocks to the application of digital tools in a democratically principled way, and propose a research agenda that directly addresses these challenges. We argue that an interdisciplinary effort integrating insights from social, economic, psychological, political and computational sciences, as well as the application of formal modelling and rigorous empirical testing, is essential to realize impactful digital democracy that can gain the trust of citizens.

\paragraph{Participation that counts}

Participation needs to be impactful in order to overcome the divide between citizens and decision makers. We use the term ``citizens'' here not in its strictly legal meaning, but rather to refer to all the people residing in a given community and having rights of participation towards the policy making of that community. The outcomes of citizen participation initiatives need to feed into the legislative process and shape political decisions. To be sure, citizen participation can inform political decision-making in various ways even without direct impact on policy making \cite{simon2017digital}. However, forms of participation that are lacking an ex ante discernible effect on policy will likely foster frustration, cynicism, and distrust in citizens and will worsen the divide between them and political decision makers. Even one of the most ambitious citizen participation events to date, the EU's Conference on the Future of Europe, has been criticized for a lack of ``clarity on the follow-up'' and ``concrete recommendations to lead to legislative action or initiation of Treaty amendments'' \cite{nguyen21}.

It is our view that a vicious circle of distrust between citizens and political decision makers constitutes the major roadblock to more impactful citizen participation. Digital tools such as communication platforms like Facebook and X (formerly Twitter) have allowed the public to express their concerns, but are also a breeding ground for opinion polarization, the spread of fake news and conspiracy theories, as well as troll attacks against decision makers, and hate speech. All of this fuels distrust on the side of political actors. While there is broad agreement on the importance of political participation, there is also growing reluctance to actually involve citizens digitally. Online communication is perceived as an unproductive black box, likely to generate digital outrage, personal harassment, even offering a point of attack for political competitors or subversion by foreign powers. In fact, before the launch of the Conference on the Future of Europe, 12 EU member states demanded that the conference ``should not create legal obligations, nor should it duplicate or unduly interfere with the established legislative processes.''\footnote{``Conference on the Future of Europe:
Common approach amongst Austria, Czech Republic, Denmark, Estonia, Finland, Ireland, Latvia, Lithuania, Malta, the Netherlands, Slovakia and Sweden'', 2021,
\url{https://www.staten-generaal.nl/eu/overig/20210325/non_paper_conference_on_the_future/document}} Alarmingly, this distrust reinforces citizens' feelings of disconnect with policy making: the reluctance of politicians to involve citizens fosters citizens’ feelings of being excluded from decision-making, which in turn further demotivates politicians to involve (disgruntled) citizens. 

\paragraph{Preconditions for impactful democratic participation}
We argue here that the dynamic of distrust outlined above can effectively be overcome if citizen participation is approached with the goal of meeting three key preconditions. 

{\em First}, if participation is meant to translate into actual policy decisions, it is critical that participation happens according to a {\bf {\em transparent}} process. How the process unfolds should be clear to participants. Steps need to be diligently documented in a privacy-respecting but traceable way, in order to be able to demonstrate that outcomes are not the result of arbitrariness or manipulation. Without proper documentation, citizens may not be able to gain trust in the process and political actors may more easily disregard the outcomes of a participatory process.

{\em Second}, it is critical that participatory processes are {\bf {\em open}} and {\bf {\em fair}}. Every citizen able and interested in participating in decision-making should have the opportunity to do so as a peer. This includes: on the one hand, providing avenues to help citizens overcome digital divides, to transcend cultural and gender norms around participation, and to enable participation in multiple languages; on the other hand, to guarantee that the participatory process satisfies key democratic desiderata of equality, inclusivity and representativity. When implemented at national or even international levels, this can imply that millions of citizens may want to contribute and that vulnerable, hard-to-reach, and marginalized groups get the voice they need. If participation has direct impact on the lives of citizens, then every citizen needs to be given the opportunity to participate. 

{\em Third}, impactful participation requires participatory processes that are {\bf {\em fit for purpose}}. Such processes should be safe from undesired dynamics like the spreading of disinformation, opinion polarization, hate speech, and be secured against fraud. At the same time, they should be able to fully reap the benefits that the diverse insights emanating from wide participation can bring. It is important that malicious behavior by trolls, bad-faith political actors, and exogenous powers is kept at bay, while the contributions of good-faith participants are capitalized on. Democracy is already experiencing a crisis. Well-intended but not well-designed participatory processes have the potential to deepen this crisis rather than strengthen trust in democracy. The public and scholarly debate about echo chambers on online social networks, about social bots, fake news propagation, and foreign interference in democratic elections illustrates how digital communication technology can contribute to undemocratic outcomes of public debates. Participation platforms need to be developed in a way that enables them to prevent such undesired dynamics {\em by design}. 


\paragraph{The untapped potential of digital democracy for impactful participation}
The current state of democratic participation is worrisome. While there are clear paths for companies and interest groups to lobby their ideas into democratic decision-making processes, citizens have very limited opportunities to participate in deliberation and decision-making. Compounding this problem, current attempts at enhancing democratic participation still fall short of delivering the means for impactful citizen participation. And this holds even for the most high-profile efforts, such as the Conference on the Future of Europe we mentioned above. Digital technology holds the promise to enable democratic participation that is transparent, open, and fit for purpose. However, while public authorities from all levels of government increasingly turn to digital democracy tools to improve citizens’ participation  \cite{OECD,simon2017digital}, the true potential of digital approaches is hardly exploited.

{\em First}, digital communication fosters transparency as it allows to document who was when contributing what to a given debate or decision and to make this information publicly available. Existing approaches to digital citizen participation are often criticized for a lack of transparency as to how the input of citizens was ultimately incorporated in the final decision-making and for including multiple filters between citizens' input and final outcome.\footnote{Criticisms of precisely this nature have again been aired against the Conference for the Future of Europe \cite{youngs22}.}

{\em Second}, digital citizen participation fosters accessibility. Online social networks demonstrate that digital communication is possible even on national and international scales. What is more, digital platforms allow users to interact asynchronously. That is, digital communication does not require that participants meet at a given time and place, alleviating one substantial burden of participation.  Existing digital citizen-participation approaches, however, often limit participation to the formulation and support of petitions and, thus, fail to support a process of genuine deliberation. Likewise, approaches that do include deliberation are often restricted to a few hundred participants usually selected at random by sortition. While this approach has proven effective in eliciting fruitful and more representative input for decision makers \cite{reybrouck16against}, its legitimacy as a tool to determine political decisions is open to objections because sortition is, in and of itself, a barrier to participation.

{\em Third}, digital citizen participation is mediated by algorithms supporting deliberation and decision making. Precisely this algorithmic layer of digital tools, once properly leveraged, makes it possible to carefully craft participatory processes that are truly fit for purpose. Existing open-source platforms such as LiquidFeedback \cite{behrens2014principles} or Polis \cite{small2021polis} (see Section 2 for an overview of the state of the art) show this is possible. In 2018, the government of Taiwan crowd-sourced legislation, using Polis, that enabled over 4000 citizens to deliberate for four weeks on how to regulate Uber’s operations in Taiwan. The results of this large-scale deliberation were adopted by the country’s legislators and became law.

While these attempts are encouraging, we need to move on from the current trial-and-error approach to the development of robust digital democracy technologies that shall serve us well in years to come. Errors in the current phase involve the failure of participatory processes: not something we can allow ourselves, given the already dire state of citizen participation. A more principled approach to technology development is needed. Crucially, such an approach is within reach. Scientific disciplines as diverse as philosophy, economics, computer science, political science, sociology, and social psychology already provide a rich theoretical and empirical toolbox to tackle the systematic design of platforms that are fit for enhancing democratic participation.

\section{No digital democracy without digital democracy science
}

For the reasons outlined above, our aim with this position paper is not to contribute to the philosophical and normative debate about what democracy is. We take a much humbler, and practice-oriented perspective and aim to contribute to the development of democratic institutions from a ‘procedural’ view-point. We view digital democracy as a library of computational tools that support such practices at various levels in our societies, from the local, to the national and the trans-national. The challenge is to identify the combination of tools that best meets the standards of transparency, openness, and fitness for purpose in the settings that are relevant for the institutions or grassroots organizations that intend to deploy such tools. 

\subsection{The state-of-the-art of digital democracy}\label{subsec:SoADD}

\paragraph{Digital participatory tools: a short history}
Technology has been part of democratic practice since at least Athenian democracy. In ancient Greece, for instance, the so-called \emph{kleroterion} was used to randomly select citizens into councils and public offices. But it was the advent of digital technology that brought the relation between technology and democracy to the forefront. Starting in the 1960s, scholars such as Gordon Tullock envisioned how emerging information technologies could be used to enable mass direct participation to the legislative process on a national level \cite{tullock67toward,tullock1992computerizing}. Similar perspectives have been echoed for decades by democratic theorists, from Robert Dahl \cite{dahl2008democracy} to, more recently, Hélène Landemore \cite{landemore2020open,bernholz2021digital}. Currently, we are faced with a wealth of platforms for democratic participation. These have been used for a variety of purposes: from citizens' assemblies \cite{parsons2019digital}, to participatory budgeting \cite{matheus2009models}, to policy-drafting \cite{randma2022engaging}, to full-fledged decision-making \cite{simon2017digital}. Prominent examples are: Polis \cite{small2021polis},\footnote{\url{https://pol.is}} which was used in 2018 by the Taiwanese government to crowdsource high-profile state-wide legislation on controversial matters such as the regulation of ride-sharing services; Consul,\footnote{\url{https://consulproject.org}} which was developed by the Municipality of Madrid and has been used by several cities worldwide and received the United Nations Public Service Award; YourPriorities,\footnote{\url{https://citizens.is/your-priorities-features-overview}} which has been used for crowd-legislation and policy-making in Iceland since 2010; Decidim,\footnote{\url{https://decidim.org}} which has been used by the city councils of Barcelona and Helsinki; LiquidFeedback \cite{behrens2014principles},\footnote{\url{https://liquidfeedback.com}} which was developed in the early 2010s and has been used by various political and civic organizations across Europe since. Many more platforms exist, as demonstrated by the online databases of Democracy Technologies\footnote{\url{https://democracy-technologies.org/database/ }} and The Democracy Foundation\footnote{\url{https://democracy.foundation/similar-projects/}}. Beyond the anecdotal successes in the deployment of participatory tools such as the ones above, evidence is slowly accumulating that technology-mediated citizen participation can improve perceived legitimacy and inclusiveness, as well as the quality of policies and government decision-making \cite{fung2015putting}. It is high time to fully tap into this potential.

\paragraph{The Conference on the Future of Europe}
The most ambitious digital participation initiative in Europe so far has arguably been the ``Conference on the Future of Europe’’, an event organized by the European Union that reached, digitally, more than 650,000 European citizens and generated a joint declaration that was endorsed by the European Parliament (resolution 2022/2648(RSP)).\footnote{
\url{https://www.europarl.europa.eu/resources/library/media/20220509RES29121/20220509RES29121.pdf}} One of the central instruments was the software {\em Decidim}, which allowed citizens to contribute in the 24 official EU languages and supported more than 52,000 active participants to debates and discussions. These contributions informed four citizen panels consisting of 200 randomly selected European volunteers. One of these panels was concerned with the ``perceived distance between the people and their elected representations’’ \cite[p. 20]{reportCFU}, the same observation that motivates the present position paper. Recommendations formulated by these and other panels were repeatedly debated by a plenary composed of citizens and politicians. The outcome of this process was a final report listing 49 proposals for reformation of regional, national, and European policy.
One of the central proposals developed by this large-scale citizen participation was that citizens demanded regular direct participation on regional, national, and the European levels. They also explicitly requested digital solutions. Similar proposals were formulated in the Charter on Youth and Democracy, another European participation initiative.\footnote{\url{https://cor.europa.eu/en/our-work/Pages/charter-youth-democracy.aspx}} In this charter, young Europeans demand more direct participation in regional, national, and international legislative processes and explicitly call for digital tools such as e-voting.

However, despite the great ambition and the impressive deployment of digital tools, the Conference on the Future of Europe has also been criticized. Most importantly, it remained unclear what the impact of the conference was. Citizens formulated their demands and the European Parliament endorsed the final report, but no clear legislative response followed. Markedly, the three aforementioned conditions of transparency, openness, and fitness for purpose were not met. There was a notable lack of transparency in the process transforming citizens' input into the final set of proposals.
Second, the most significant part of the deliberative process and the actual decision-making were restricted to a very small group of citizens (panels and the plenum). For most citizens, participation was limited to writing and endorsing comments. Third, it is unclear to what degree the participation process was actually fit for purpose. Can we consider the outcome to really represent the demands of European citizens? Have the main issues citizens consider important been covered? To what degree has the outcome been shaped by the politicians and experts who have been involved in the process? These weaknesses limit the impact of this ambitious and visionary citizen participation process, precisely because it is all too easy to disregard its outcomes as failing to truly reflect the interests of citizens. 

In a nutshell, the Conference on the Future of Europe serves as an ideal-typical illustration of the currently prevailing approach to digital participation. On the one hand, the conference demonstrates that large-scale digital participation is technically possible and that there is a strong demand on the side of citizens. On the other hand, it also illustrates that designing digital participation platforms is challenging and that problematic design decisions may generate frustration and potentially even intensify the sidelining of citizens.

\subsection{The core technological challenge: from democratic values to software}

\paragraph{Democratic values in digital democracy}
We documented above a strong demand for more citizen participation in general, and for digital means to enable such participation in particular.
Any digital tool for democratic decision-making and participation operates in a context that is necessarily value-laden. There are general fundamental values that both conceptually and legally define democracies (participation, autonomy, rule of law) and fundamental rights and goods (free speech, various freedoms from intervention and harm, protection of minorities, as captured, for example, in fundamental human rights). These values, in practice, are subject to trade-offs, both at the abstract level and in the individual case (e.g., effectively protecting individuals from harm may require prohibiting certain kinds of speech). Those tensions and trade-offs are fundamentally not ours to resolve, last but not least because they will be resolved differently across countries and across contexts. At the same time, it is important to acknowledge that technology is not value-neutral \cite{van2011ethics}, and this is even more so the case in the democratic context we want to address.


\paragraph{From values to software}
Concrete system design decisions will impact features such as effective participation and will likely do so in multiple complex ways. Importantly, designers will also have to be vigilant to prevent discriminatory biases and include the necessary safeguards. This means that we must necessarily adopt an \emph{instrumental perspective}: \emph{If} we want democratic tools to satisfy (or even promote) a particular system behavior, \emph{that} is how the software should be designed. Even with that framing, there will be multiple limitations. Once understood what features promote impactful participation, there will still remain instrumental questions about how manipulating those features in a digital system promote the wider democratic goals. 
Such a question may sound simple, but the answer will likely be the result of multiple independent, or semi-independent components, design features and their interactions. Coming to understand these components and design features, and how they jointly produce particular outcomes is what we believe is currently missing for the development of fit-for-purpose digital democracy solutions. {\em Understanding how specific design choices affect the democratic quality of digital democracy systems is the focus of the research efforts we are promoting}.

\paragraph{Digital democracy as a complex socio-technical system}
We see it as a likely consequence of the nature of the type of systems under study---complex systems of multiple interacting individual agents with the possibility of feedback dynamics---that there will be no single unique answer of how to design for effective participation. Instead, the likely outcome of research in this area will involve formulating a space of possible designs along with evidence-based guidance on what designs might be more or less appropriate with respect to particular value-based choices in a given context. Despite the huge interest and demand in digital democracy, the interdisciplinary research endeavour aimed at these tools is presently still rather far from such a deep instrumental understanding of the design space of tools. Progressing toward that goal, given its intrinsic complexity, will consequently have to resort to a rich combination of methods.

\subsection{The core research challenge: democracy requires scientific rigor}

As argued in the previous section, design decisions have an impact on whether and to what degree a given software is able to realize democratic values. Thus, it is critical that such decisions are made explicit and that they are explained. That is, developers need to be able to demonstrate that a given design contributes to reaching a given democratic value. This level of confidence in digital tools cannot possibly be achieved by trial-and-error. It requires a valid theoretical argument about why a given design decision has a desired consequence, as well as the empirical research necessary to empirically support the theoretical argument.  

Fortunately, various scientific disciplines provide a rich toolbox of theoretical models, formal methods, and empirical approaches that have proven effective in rigorously developing scientific understanding of complex social systems and computational artifacts. We argue that these tools need to be applied to the development of digital democracy in order to rigorously anticipate the consequences of design decisions and to best implement democratic values in software. Below, we outline the three methodological pillars of the scientific approach to digital democracy we argue for in this position paper: modeling, behavioral experiments, observational studies.

\paragraph{Why formal and computational models?} 
First, we argue that mathematical and computational models are key for the development of digital tools for digital democratic participation. Rather than developing theoretical arguments based on intuition, formal approaches use logic, mathematics, and computational methods to develop valid theory. In particular when the system under investigation is complex, formal modeling allows one to test the logical validity of one's intuition and to rigorously demonstrate that a given theoretical argument is valid \cite{liu2022intuition,schelling2006micromotives,epstein2008model}.
This theoretical transparency is a value in and of itself, especially in the context of the development of democratic tools. Furthermore, mathematical analysis allows one to explore in detail the extent to which results rest on any one of their assumptions. In this way, formal methods allow one to move beyond intuition to a fine-grained understanding of a problem space. This is especially useful for the analysis of the algorithms powering digital democracy systems \cite{brandt2016handbook}.
But besides these general methodological points, there are three more concrete functions of formal methods that specifically match our purposes.

{\em First}, considering that democracy is already under pressure, we deem it critical to deploy digital tools only when one is confident that they satisfy core democratic values and do not generate undesired effects. The public and scholarly debates about filter bubbles, social bots, and fake news on online social networks illustrate that seemingly innocent design decisions in digital tools can have highly unexpected yet impactful consequences. These issues will likely be amplified further by the ever faster inclusion of artificial intelligence components in more and more areas of life and the fact that the behavior of existing AI systems is typically not well understood, let alone formally verifiable. Formal methods that are carefully tailored to the application contexts they represent, on the other hand, enable one to demonstrate mathematically how envisioned design satisfies democratic criteria, such as fairness, equity, and representation. Furthermore, models that are carefully calibrated to the application domain they represent make it possible to gain insights into digital democracy tools before they are actually deployed in order to anticipate undesired effects. In other words, formal methods make it possible to demonstrate to citizens and decision makers that digital tools meet the democratic desiderata expected from them even before they are actually deployed. 

{\em Second}, approaches to digital democracy that rely exclusively on real-life experimentation can be fundamentally limited. A central problem is that experiments with digital tools may fail and further damage citizens' trust in democratic institutions, in particular when implemented on a large scale. Unfortunately, there are also limits to the gradual up-scaling of digital participation. As a consequence of this, it is certainly wise to experiment with digital tools only on a small scale. However, “more is different” \cite{anderson1972more}, a notion that can apply to any system that consists of micro-entities exerting influence on each other. Even within the field of physics, for instance, Anderson famously argued that the “behavior of large and complex aggregates of elementary particles [\ldots] is not to be understood in terms of a simple extrapolation of the properties of a few particles. Instead, at each level of complexity entirely new properties appear” \cite{anderson1972more}. 
This echoes an argument in the social sciences developed already by Emile Durkheim, one of the founding fathers of sociology, who reasoned that “society is not the mere sum of individuals, but the system formed by their association represents a specific reality which has its own characteristics” \cite{durkheim2016rules}. 
Thus, drawing conclusions about the dynamics in large populations based on observations made in smaller communities is inherently problematic. Computational models enable extensive experimentation at scale and provide insights into the behavior of digital democracy tools in large communities that would not be accessible via observations alone.

{\em Third}, and related to the above point, formal methods allow one to study counterfactuals and to systematically improve digital tools without experimenting in real life. A formal model that has been carefully calibrated to a specific empirical setting can function as a so-called ``digital twin'', a computational model that mimics all relevant aspects of the real system it is supposed to represent \cite{tao2019make}. NASA, for instance, developed digital twins of their space vehicles since they cannot be investigated once deployed in space. What is more, copies on Earth are of limited use, since they are not exposed to the harsh environment of space. Digital models of their vehicles, however allow one to simulate how they cope for instance with radiation. Thus, digital twins can be used to predict dynamics of a system without studying the real system. Moreover, aspects that cannot be changed in the real system can be manipulated in the digital twin without ethical, technological, or financial limits. Comparing the digital twin with these counter-factual siblings allows one to rigorously quantify differences between the two models. In the context of digital democracy tools, for instance, one could use a digital twin that has been carefully calibrated to deliberation dynamics in small groups and study dynamics in much bigger systems with millions of virtual users. AI systems, in particular large language models, can be of help here (e.g., \cite{bakker2022finetuning,fish2023generative}), but their use in scientific studies also poses novel challenges of explainability and reproducibility.

\paragraph{Why behavioral experiments?}

At the heart of research on human behavior is the ability to run behavioral experiments, that is, studies in which factors are \emph{manipulated} across experimental conditions in a study with participants typically drawn by random sampling, so as to afford generalization of results to a wider population. 
Experimental manipulation of putatively relevant factors is what allows one to infer causal effects, whose presence or absence (and effect sizes) are typically ascertained through the use of inferential statistics. 

Behavioral experiments vary with respect to how precisely individual `factors' are isolated and how `random' the sampling of the study sample is. At one end of the spectrum, laboratory studies allow careful stimulus control. Thus, in conjunction with judicious sampling of participants, laboratory studies allow one to eliminate confounding variables. However, this degree of control may come at the expense of so-called `ecological validity', that is, the extent to which the laboratory task still matches the conditions experienced `in the wild'. This may limit the transferability of results obtained under tightly controlled laboratory conditions to real world settings. At the other end of the spectrum, field studies conduct experimental manipulations `in the wild', with limitations on experimental control and sampling. There is, arguably, no such thing as a `perfect experiment' or a `perfect experimental design', as a result. Rather, results from behavioral experiments are strongest where they can draw on convergence across methods. 

Importantly, the digital environments of behavioral experiments can conveniently reflect real world `digital environments' to a greater extent than is possible for many other domains of inquiry, simply because digital tools can be used to simulate the very environments of interest even where those environments are not actually present.
At the same time, experimental conditions can be created to match most closely either the real world or a particular model chosen to study the system (more on this below). In this context, behavioral experiments serve to connect micro and macro behaviors of a digital system.

A further feature of behavioral experiments that is utilized particularly within economics and behavioral economics is the possibility of incentives. Psychologists and economists disagree somewhat on the need for providing explicit incentives for participants in behavioral experiments \cite{read2005monetary}. Regardless of how one comes down on the issue of whether behavioral experiments should be incentivized as a matter of course or not, incentives offer an important additional tool: incentives make decisions meaningful by including utilities, and this allows one to study preferences and the intensity of those preferences. Incentivized experiments allow researchers to measure and experimentally manipulate individuals' preferences with great precision. Specifically,  when a behavior is made costly but is engaged in nevertheless, there is strong evidence that there is a motive. Furthermore, by varying those costs, one can determine how strong the motive is. In the context of building a science of digital democracy, this means, for example, that one can measure the degree to which individuals consider certain decision-making processes unfair. Yet, people are motivated by different things: financial incentives, status, reputation, altruism, higher meaning, religion, ideology, loyalty, and so forth. How much does habituation matter? Could children be socialized into lifelong democratic participation by engaging (and empowering) them early on in matters that affect their lives? Beyond families, could schools, as a form of civics education, involve children in context-appropriate decision-making and instruct them in idea formulation, deliberation, decision-making, and voting? Experimental designs in digital democracy should incorporate awareness of these factors.


One important difficulty, finally, for the context of an empirical science of digital democracy is that fully understanding human behavior in a collective may also require experiments with \emph{groups}. Such experiments are vastly more costly, both from a logistical and a resource perspective, than experiments involving individual participants alone. In particular, what from the perspective of statistical analysis was an individual, is now replaced by a group. Given that the need for statistical power (which characterizes sample sizes required to detect a given level of empirical effect) will often require sample sizes that place a considerable financial burden on experimenters (for participant payments and incentives), theoretically desired experiments with groups may often not be realisable in practice. This is exacerbated by the possibility of `more is different' phenomena highlighted above, whereby studying a small to medium-sized group may not all translate readily to the dynamics of larger groups. 

Coupling simulations with experimental studies to derive suitable `digital twins' is one way of addressing these challenges. 
Another path forward may lie in the future creation of shared large scale resources, a kind of CERN for the study of information environments, that would involve the creation of online platforms that are available for experimental testing \cite{lewandowsky2020technology,CERN}.
Yet another alternative is to draw on the complementary strengths of non-experimental, observational, methods.

\paragraph{Why observational studies?}

Observational studies lack the defining features of experiments: an experimental manipulation. This makes them less suited to identifying causal effects, although statistical methods exist to aid causal inference from observational data \cite{breen2022causal}. This limitation is compensated for by the richness that qualitative data can provide, including its typically greater external validity: observing a behavior in a real-world context does not suffer the same problems of distortion that the development of a tightly controlled experimental stimulus may entail. Consequently, observational studies may help identify and understand limitations of behavioral experiments and simulations via computational models.

Observational studies, too, come in many forms and may be either qualitative or quantitative. Among the latter, the analysis of social media data is a natural element for the study of digital democracy. Here, the burgeoning field of computational social science \cite{flache2022computational} has provided many studies seeking to understand how and why information spreads in online communities (see e.g., \cite{vosoughi2018spread}, but see also \cite{burton2021reconsidering} on the methodological difficulties). Furthermore, such data may connect quite readily with measures used to evaluate computational simulations.
At the same time, it is important to remember the limitations inherent in the samples that are available for study by researchers, although recent legislative developments such as the EU's Digital Services Act have also sought to widen the accessibility to researchers of data obtained by large social media companies \cite{nonnecke2022eu}.

\medskip
\noindent
We end by stressing that we ultimately not only need to use all the above methods, but that we should think about how they can be most effectively deployed in concert. That is an important methodological aspect of the research agenda that needs to be developed, albeit one which is not pursued further in this paper. 

\section{Towards a research agenda}

In what follows we outline, in some detail, six research challenges that we consider particularly salient. These challenges have been defined by working groups at the Lorentz Center Workshop on ``Algorithmic Technology for Democracy'', which prompted this position paper. Such challenges constitute a starting point, and by no means an exhaustive list. At the same time, they exemplify well the complexities of digital democracy research that we identified above. 


\subsection{Equality in digital democracy}

Equality may be considered one of the \emph{raisons d'être} of democracy. In this view, democracy is not an end in itself, but rather a means for people to have equal standing in all aspects of the political decision-making process. Accordingly,  the success of a process---including a digital one---in providing such equality should be one of the key measures by which it is evaluated, critiqued, and corrected.
Traditional democracies have practical limitations on granting equality throughout the entire political process, hence they typically make do by providing citizens with the right to elect their representatives every few years. Some democracies provide also citizens with due process to initiate new legislation, including amendments to the constitution, as well to annul any legislation passed by their elected representatives---the parliament. As already observed by political scientists in the 60s \cite{miller1969program,tullock67toward}, digital democracy has no such practical limitations, and therefore may open for investigation all aspects and components of the political decision process regarding whether and how they respect equality when being conducted in the digital realm.
For concreteness, we list some key such components of the decision-making process below:
\paragraph{Equality in agenda setting.}  Decisions on what proposals to vote upon, and in what order, may have tremendous impact on the final outcome of the political process.  {\em How can the power to set the agenda and to make proposals be shared equally among all participants?} \cite{bulteau2021aggregation}

\paragraph{Equality in deliberation and coalition formation.} Similarly, processes that precede a binding vote can also have tremendous impact on the final outcome.  {\em How can deliberative processes}, which may include negotiations on amendments to proposals to be voted upon and the formation of coalitions behind  proposals or behind amendments to existing proposals, {\em be structured in an egalitarian way, giving all participants an equal footing in the process}? \cite{elkind2021united}



\paragraph{Equality in decision making.}  How can we ensure one person -- one vote in the digital realm? Fake and duplicate digital identities (aka Sybils), coercion, bribery, and fraud are all possible also in the physical realm, but their risk is amplified in the digital realm~\cite{meir2022sybil}.  In this context, a crucial issue for the development of digital democracy is to understand how digital participatory processes should interact with standard institutions of representative democracy. One may also ask which votes, or decisions, should be taken via direct open digital processes, and which ones via the standard representative institutions: {\em what should be the division of labor between digital democracy and representative democracy}?
    

\subsection{The rule-tool-user nexus in digital democracy}


Collective decision making is the main building block of democracy. In a process of collective decision making, a group of people (or \emph{agents}) interact by exchanging arguments or by expressing opinions about some issues at stake, and they submit certain data to a mechanism, e.g., by filling in a ballot. A final decision is usually produced by \emph{tallying}---transforming this information into a result via a voting \emph{rule}, e.g. by aggregating ballots. In cases where the decision-making process is supported by a platform, agents act via their \emph{user} profiles, while deliberation and tallying are carried out through a voting \emph{tool}.
The three main components of digital collective decisions---the \emph{users}, the \emph{rule}, and the \emph{tool}---are not independent \cite{terzopoulou2023rule}. Their behavior relies on the existence of each other: they anticipate feedback and (should) respond to it.
As in every \emph{complex system}, governing forces are the feedback mechanisms and not the particular elements of the system in isolation. Importantly, although agents that form opinions and make decisions together within groups take part in a complex system as well, social influence and collective decision making have so far been studied separately. By treating digital collective decisions as a complex system, we improve our understanding of their properties and capabilities, and enable further progress on the platforms where they take place. 

\paragraph{User--user interaction} 
Digital collective decisions are not simply the result of voting rules applied to some fixed ballots submitted by the users of a platform. Users often behave strategically, deliberate, and influence each other; their preferences that lead to the ballot formation are subject to complex processes of social dynamics, which are magnified in digital contexts and may severely bias the outcome. Online anonymity facilitates the expression of extreme opinions and sustains polarization, while social networks entail the risk of information bubbles that damage diversity of information diets \cite{keijzer2022complex,flaxman2016filter}. When 
collective decisions are at stake, imitation effects might increase the predictability of results due to easy consensus, or decrease it due to multistability or oscillatory behavior \cite{torok2013opinions,lorenz2018opinion,ganser2018social,flache2017models,BaraLT22}.  Many relevant research questions arise: How should a digital voting process be designed to benefit from positive social-influence effects and mitigate negative ones? Are collective decisions vulnerable to cross-platform interaction of the users? How can digital platforms for collective decisions encourage information-seekers and disincentivize fake news? How do we identify and prevent vote manipulation, in groups with strong tendencies to either grow together or grow apart?

\paragraph{User--rule interaction} 
As long as the goal of digital platforms for collective decisions is to support the open, equal, and transparent participation of everyone, the voting rules employed must ideally account for user heterogeneity. Users may differ in terms of their \emph{strategic} or \emph{collaborative} behavior, \emph{motivation}, and their \emph{perception of axioms}. This should be considered when trying to build rules that are hard to manipulate, and rules that provide incentives for participation. In social choice work to date, well-known impossibility results rely on strong assumptions of homogeneity \cite{campbell2002impossibility}, but in digital contexts assumptions about a group of users should be directly linked to the platform in which the decisions take place: for example, a platform that requires several steps for registration will naturally attract more motivated users, while a platform that allows a user to create multiple accounts may increase strategic behavior. Most importantly, the relation between a voting rule and a group of users is usually considered one-way: the users provide the input to the rule. But the other direction of the relation is also important: the rule can affect the users, steering them towards a specific type of ballots and preferences 
 \cite{van2013vote} (for instance, a rule that asks for approval ballots will more likely incite binary preferences than a rule that asks for complete rankings).   
A rule may be perceived by a user through three lenses: the \emph{axioms} it satisfies (i.e., reflection of the user requirements in the properties \cite{procaccia2019axioms}), the \emph{procedure} it follows (i.e., cognitive simplicity and explainability of the procedure \cite{BoixelEtAlIJCAI2022}), or the \emph{outcome} it produces. Examining closely the interaction between users and rule highlights several research gaps, on both the theoretical and the empirical front: What are the potentially heterogeneous attributes of users? What are the multiple values or forms they may take? Which of them could possibly be influenced by the voting context? How does the conceptual goal of a voting rule relate with its mathematical definition?

\paragraph{User--tool--rule interaction} 
In the digital world, voting rules are implemented in the form of tools. At the least, these tools let users set up decision problems, list options, and fill in their secret ballots, before performing the tallying and reporting the results. A tool might offer several different ways for filling ballots, e.g., specifying an integer rank for each option or dragging options up and down to specify a ranking. For voting rules where an effective use of one's ballot depends on beliefs about others, the tool might help in forming these beliefs, e.g., by providing voters with voting data  and letting them adjust their ballot once, several times, or as often as necessary until some deadline. Software tools can be designed to provide quite varying user experiences \cite{hassenzahl2006user}. 
The choice of language and terminology and the overall visual design (colors, fonts, imagery, etc.) might induce framing effects and prime user behavior in certain ways, e.g., to be more cooperative or competitive, individual- or social-value oriented, honest or strategic, process- or outcome-oriented, etc.
\cite{capraro2019power,gong2013framing,herr1986consequences,drouvelis2015can}.
Design details during voting or deliberation such as the order of questions, hints or nudges, or information on other users' behavior can influence users in wanted, unwanted, or unpredictable ways \cite{krosnick1987evaluation,sunstein2014nudging,capraro2019increasing}.
Being the main interface between users and voting rules, voting tools need to be understood and designed properly, to assess and improve the impact of digital democracy. We locate a number of imperative research questions:
How can or should a certain formal rule be turned into a tool? Which of potentially several equivalent representations of its input, algorithm, and output shall be chosen? 
How can usage data from a tool be exploited to assess quantitative formal properties of a voting rule (such as welfare or satisfaction metrics, frequencies of axiom violations or strategic behavior, degree of engagement, etc.) and test related theoretical claims? How can data from tools implementing different rules be used to advance theory (e.g., by suggesting additional formal criteria, ballot designs, or more complex game-theoretical models of a decision rule in the  context of certain interactions)?


\subsection{Citizens as users: preference elicitation in digital democracy}

Understanding citizens' preferences is arguably one of the primary goals of any digital democracy platform.
Elicitation is the process of extracting or producing information, often in reaction to a query or question. When applied to preferences, such a process includes preference formation, since users may not be aware of their preferences before being asked.
Preference elicitation is the point of contact between the users and the democratic procedures they engage with. The quality and quantity of the collected preference data---and of the democratic process following---heavily depend on such points of contact. 
One should not be surprised then to see a vast literature on elicitation in closely related fields such as voting theory and computational social choice, multi-criteria decision analysis, recommender systems, and automated negotiation \cite{Lee2014,mousseau2015preference,benade2021preference,BaarslagGerding2015,GARCIA2012155}.

\paragraph{Formalizing preference elicitation protocols} Elicitation is supposed to track users' preferences on a set of proposals (or alternatives, or candidates), which may be described by features. Users can also have beliefs on other users’ preferences, and may be not be aware of the existence of all proposals.
An elicitation protocol is a sequence of elicitation actions that extracts information from users and represents it in a computationally-friendly format, designed towards the solution of a collective decision problem.
%
Elicitation actions can take the form of explicit queries on preferences, e.g., asking users to specify which is their top-preferred proposal, approving a set of proposals, expressing pairwise comparisons.
Other elicitation actions are: collecting preferences at the level of features of the proposals; eliciting awareness or belief about other users’ preferences; asking whether given proposals share certain features; asking users about the expected benefit of a proposal.

However, a fundamental problem of preference elicitation via any form of communication is that an individual's actual preferences are mental states that are not directly measurable, and in most contexts where preferences are elicited, individuals might have incentives to misrepresent their preferences, either due to strategic interaction with other individuals' preferences via some form of collective choice mechanism, or due to social-psychological effects related to social image, expectations about desired behavior, fears of punishment, etc. As a consequence, stated preferences might differ systematically and in complex ways from actual preferences. 

\paragraph{Research challenges} The use of digital tools in democracy allows for collective choices over large and possibly complex sets of alternatives.
Elicitation protocols thus need to take into consideration the \textit{high computational and communication costs on the protocol side}, but also constrain the \textit{cognitive load and the time effort on the users' side}.
Another related issue is the handling of incomplete preferences, which may occur due to several reasons, such as the high cognitive load required for submitting complete preferences but also due to other factors such as general imperfect knowledge.
Transfer and inference of preferences, either in time or between individuals, might be a viable solution here. 
Also, ethical considerations must be taken into account, such as biases in the elicitation protocol and in the inference of preferences.

\subsection{Political representation and digital democracy}

%

Representation is a core democratic value \cite{mill1924representative,mansbridge2009selection,landa2020representative,rae1967political,lijphart1990political,michener2021remoteness} but it is yet to be understood in a more detailed way to be operationalized in digital tools. Whatever form digital democracy takes, it will need to be able to credibly claim to represent the citizens it serves and to work along traditional, as well as novel, institutions of political representation. 

\paragraph{Innovation in representation mechanisms} There is no shortage of innovative proposals to change how representative bodies are selected around the world. 
For example, some propose
to select representatives at random (a.k.a.\ \textit{sortition}) \cite{bouricius2020democracy},  to elect them through transitive delegations (a.k.a. \textit{liquid democracy}) \cite{blum2016liquid,valsangiacomo2021political}, or to drastically increase the size of parliaments (see, e.g., \url{https://thirty-thousand.org}). Each proposed method has its benefits and drawbacks; however, we lack a systematic way to evaluate and compare them.
Specifically, while there are numerous works in computer science and political science analyzing the strengths and weaknesses of specific methods, principled comparisons are rare.%
\footnote{The few existing exceptions mostly focus on epistemic aspects, the robustness of representation, and majority agreement \cite{alouf2022should,Gree15a,gelauffrepresentational,abramowitz2019proxy}.}

\paragraph{Research challenges} We call for the development of a unified framework to formulate and compare innovative and traditional mechanisms for selecting representative bodies on a more principled basis. 
Having formulated different mechanisms within the same framework then offers the possibility to formulate various desiderata (that is, democratic values) in the same framework.
While we believe that comparisons from different perspectives are possible and, in fact, urgently needed, we argue for the development of an \textit{axiomatic} view on selection mechanisms, drawing inspiration from the rich social choice literature on voting rules.
Notably, the focus of such an approach would not be on finding the ``ideal'' representation system. We rather envision building a navigator that maps selection mechanisms to axioms. 
We advocate for building a coherent picture of the advantages and disadvantages of competing selection proposals based on carefully crafted axioms to gear public debates towards \textit{what kind of trade-offs societies are facing}, instead of continuing to argue for competing selection mechanisms on disconnected grounds. First steps on this line of research have been presented in \cite{revel2023selecting}.

\subsection{Democratic deliberation and digital democracy
}


Digital democracy has the potential to involve large parts of a population or organization in rich processes of deliberation that go way beyond online petitions or voting. As contemporary online social media demonstrate, however, digital discussion does not necessarily promote democracy and can even foster discrimination, hate speech, and the spread of disinformation.
Designing platforms that can truly foster democratic deliberation is a formidable challenge that requires substantial research efforts to succeed at scale. 

\paragraph{From e-petitions to online deliberation and decision-making}
Digital democracy can grow to include rich dimensions of deliberative interaction: from agenda setting to opinion formation to decision-making. Many platforms aim to facilitate deliberation, while others take a voting only approach~\cite{simon2017digital}. Deliberation fosters an understanding of the complex nature of a given issue, and contributes to informed decision-making. However, all too often, the opportunity to reach a decision is missed, and deliberation remains inconsequential. That's why some platforms opt for a combination of deliberation and voting. It should be noted that the result of the vote need not be a binding decision, but may also have an impact as a collective recommendation of the electorate. This requires research on how to legitimately embed credible citizen participation into the representative system.
Similar to the voting systems discussed in the previous section on political representation, different voting systems can be deployed to realize decision-making on the issues at stake: from single-winner (selecting one alternative out of many) to multi-winner (selecting some alternatives out of many~\cite{faliszewski2017multiwinner}).
When multiple winners are selected, quite different goals may be pursued. Lackner and Skowron \cite{faliszewski2017multiwinner} mention three principles: `individual excellence', `proportionality', and `diversity'. Voting systems and preference aggregation algorithms in general have a variety of additional properties that may make them especially suitable for use in specific contexts, but less so in others. A principled understanding of how to map the properties of these voting systems to their envisioned context of use is an ongoing challenge of great relevance for the development of digital democracy platforms that can support large scale deliberation and decision-making. These issues become even more pressing when considering decision making in richer settings such as, e.g., participatory budgeting.


\paragraph{Algorithmically-supported deliberation}

Scaling up deliberation to large groups interacting online and asynchronously requires algorithmic solutions to several problems \cite{velikanov2017mass,behrens2022implementation,Shortall-2022-FPS-Deliberation,landemore2022can,mikhaylovskaya2024enhancing}. We mention two such problems to illustrate the point. First, views from participants, and reactions to other participants' views, should be collected. Many examples exist. For example, the Deliberatorium \cite{klein2011} is a web-based system that allows people to share ideas, to add supporting and attacking arguments for ideas, and to vote on these. It combines ideas from argumentation theory and social computing to help large numbers of people, distributed in space and time, combine their insights to find well-founded solutions for complex problems. Similarly, one may combine ideation and negotiation \cite{fujita2017enabling}. For an overview, see \cite{shortall2022reason}. However, in a large deliberation, no user can possibly comment on all views expressed by others, so the problem arises of what views to present to each user (sometimes called `opinion routing' problem \cite{small2021polis}). 
Second, users need to be presented with some form of real-time overview of the state of the deliberation.
Here, voting algorithms with diversity or proportionality guarantees can again play an important role as equitable mechanisms for information-processing \cite{behrens2014principles,skowron17proportional,halpern23proportionality}. For both the above problems the question arises of {\em how participants can be certain that the algorithms deployed in the process respond to democratic desiderata} \cite{pool2013translating}: are all views and participants given equitable representations? Similarly, transitive proxies, i.e., liquid democracy, can be used as a mechanism to support large-scale democratic deliberations. LiquidFeedback, for example, uses transitive proxies as an empowerment for debate during the structured deliberation, to faciliate collective moderation, to determine the viable voting options, and for the final decision-making.



\paragraph{Modeling and measuring deliberation} 

Work in deliberative democracy initiated in philosophy and the political sciences in the 1970s (e.g., \cite{rawls71theory,habermas2004discourse}) and strengthened in the 1990s (e.g., \cite{dryzek00deliberative}). But it is in recent years that it found its way into practice by inspiring several democratic experiments such as citizens assemblies, which we already mentioned earlier \cite{reybrouck16against,fishkin2021deliberation}.
Unlike for voting, however, we still lack a principled understanding of how deliberative processes should be designed \cite{knight97what,pivato19realizing}, especially online. We even lack a basic toolbox of how to approach this kind of design problem, because we still lack suitable models of deliberative processes. Models exist that address specific features of deliberative processes, e.g.,: opinion change, consensus, meta-consensus \cite{list2013deliberation,elkind2021united}; 
issue identification and clarification \cite{list2013deliberation,list2018democratic};
decision-making improvement \cite{pivato2017epistemic,ding2021deliberation};
agenda setting \cite{rey2020shortlisting};
information exchange and persuasion \cite{glazer2008study,chingoma23tale}. However, {\em no sufficiently comprehensive model exists to date that could provide a basis for design or validation efforts}. Similarly, {\em no consolidated method exists to date for assessing the quality of deliberation} \cite{handbook_delib}. Some metrics have been proposed both in the deliberative democracy (e.g., Discourse Quality Index~\cite{DQI} or the Deliberative Reasons Index~\cite{niemeyer2022deliberative}) and computer science literature~\cite{salah13extracting}, but they are still either too abstract or hard to deploy in practice at scale.

\paragraph{Social psychology of deliberating groups 
}
Cooperation among individuals and a certain level of adherence to norms are vital for a group to work with mutual advantage of their members and, ultimately, for a society to thrive. The convention by which everybody should drive on the same side of the road is beneficial for everybody (although neither right-hand nor left-hand drive is intrinsically good or bad). Very often, spontaneous coordination emerges even in absence of written rules or external constraints \cite{homans2017human}, and may lead to results that outclass individual efforts, as witnessed by many instances of the so-called \emph{wisdom of crowds} phenomenon \cite{surowiecki2005wisdom}. 
At the same time, social sciences witness a host of examples of suboptimal or even catastrophic coordination, where failure is to be ascribed, quite to the contrary, to the presence of the group. Students may avoid asking a question in public not to be perceived as the one who did not understand, while clarifications would be beneficial for everybody \cite{katz1931students}. A desire for acceptance, belonging, or to prove one's loyalty may equally undermine individually rational responses. Such a collective blackout, also labelled as \textit{pluralistic ignorance}, is in fact omnipresent in daily life \cite{j1986discovery}. People often adapt to the behavior of other group members in a way that results in so-called \emph{informational cascades}, where people follow each other in line with the first one heading to disaster. This type of behavior may have effects at least as big as the bursting of market bubbles \cite{easley2010networks}. 

Folklore explanations for such problematic group dynamics are that humans are inherently irrational or at least very bounded. However, formal approaches to the study of these phenomena have shown that bad informational cascades may occur even among individuals that reason in the best possible way given the circumstances \cite{bikhchandani1992theory}. The same holds as well for pluralistic ignorance \cite{proietti2014ddl} and many other detrimental group behaviors, the occurrence of which is in fact consistent with assuming that individuals are rational. Cooperation and cohesion among certain groups in diverse societies is not a given \cite{enos_gidron_2018}. Political, ethnic, and racial group-based hierarchies coupled with discriminatory attitudes and a preference for excluding other groups from power leads to decreased cooperation and hampers cooperation \cite{enos_gidron_2018, habyarimana_humphreys_posner_weinstein_2007}. 

All detrimental group phenomena mentioned above are in some measure due to a lack of respect, trust, and communication among individuals in specific situations. As a consequence, one may think that this is the cause of trouble: once barriers are down, then everything should work. Unfortunately, this is not quite the case. Indeed, collective disasters may occur also because of abundance of information exchanged among individuals. This holds for many instances of groupthink \cite{janis1983groupthink} and group polarization \cite{stoner1961comparison, moscovici1969group}. In such cases, it is highly possible that initial errors get magnified by a dynamics similar to informational cascades, but this is certainly not due to lack of information circulating. In general, the \emph{quantity} of information circulating is a relevant factor in the emergence of both good and bad opinion dynamics. Yet, how the two things are related is highly context-dependent, and easy generalizations such as “lack of exchange = bad dynamics” cannot be easily made. 


\paragraph{Social psychology for digital deliberation design} Considerations like the above are even more relevant in the context of designing tools for digital democracy. Here, the question is not simply about designing a performance-optimizing architecture: an additional issue is that the design of the platform should respect specific desiderata such as {\em transparency} and {\em openness}, which we mentioned earlier. 


Furthermore, a large body of knowledge provided by psychologists, social scientists and behavioral economists needs to be taken into account and framed against a specific normative constraint for design. This holds not only for the case of openness and the quantity of information, but as well for other desiderata and dimensions. The \emph{structure of the network} of communication in which individuals interact may heavily influence opinion dynamics  \cite{barabasi2016network, newman2018networks}. In many cases limited connectedness among agents may even enhance the performance of a group and limit the effect of detrimental dynamics. Again, this fact needs to be taken into account when planning a mechanism where \emph{diversity} is a normative constraint, i.e., the possibility for everyone to access alternative views. Social psychology also witnesses many instances of \emph{overthinking} and the \emph{paradox of choice} \cite{schwartz2004paradox}, where evidence shows that choices among too many alternatives may affect the consumers’ preferences and induce suboptimal choices \cite{chernev2003more}. In the context of discussion about digital democracy, this suggests that also the normative ideal of \emph{representativity} has problematic aspects that need to be taken into account. In general, the message is that design guided by naïve optimism can have strong side effects and that normative ideals need to be carefully balanced against psychological realism. 

\subsection{Digital democracy, identities and grassroots digital communities}

In this last subsection we point to a key implementation issue of digital democracy: digital identities. The more our democracies rely on digital tools, the more crucial the provision and management of the digital identities of citizens become. When one has a unique digital identifier, how much one wants to reveal about oneself might depend highly on a given context and particular goals of the digital democracy applications one engages with \cite{Przybylska:2017}.  There is, however, to this date only very little research on the very notion of digital identities and on the tools that could facilitate and maintain them in a democratic way. Therefore, the research question that this problem raises is:  How could we create and manage digital identities in a democratic way?  What are technological hurdles in doing so, and where can technology enable such identity creation?

The core underlying problem lies in the nature of digital identities and their proneness to being copied or even invented from scratch. If, depending on one’s computational resources, one can duplicate one’s opinions or votes at will, then there always needs to be some external authority or institution that validates identities, checks for duplicates and filters them out. State-supported digital identities to access public services are by now supported and used worldwide and have developed further in response to the challenges posed by the COVID-19 pandemic. Whether and how such identifiers could or should be used in the democratic setting (e.g., even just for electronic voting) is an important matter of study as the legitimacy of any decision that relies on digital identities will necessarily depend on the trustworthiness of those identities.

\medskip

Alternatively to the institution-based approach to the validation of digital identities, research has focused also on how to provide trustworthy digital identities in a decentralized fashion and for grassroots digital communities: a bottom-up approach to the provision of digital identities that would seem to be impossible from the get-go. In particular, research on sybil-resilient social choice \cite{Shahafetal:2019,Tranetal:2009} and trust-based communities \cite{Poupkoetal:2019,Tranetal:2009} suggests that digital identities in grassroots communities could be based on a system of trust among community’s participants. For example, existing members of a digital community could personally vouch for incoming participants, confirming their identity and the fact that the person is ``real’’. Said approach would ensure that acceptance of new members is based on mutual trust without rules that need to be enforced by a centralized authority. This would, in turn, facilitate the self-organization of communities \cite{poupko2021building} and support egalitarian decision making including decisions concerning community management \cite{shahaf2020genuine} and its fundamental rules of conduct, or constitution \cite{abramowitz2021beginning}. This type of digital identities blueprint would thus be especially suitable for digital democracy applications implemented by grassroots civic communities. Such a bottom-up, trust-based approach may resonate with otherwise disenfranchised citizens by facilitating the self-organization of digital civic participation.  
\medskip

It should be clear that at the moment, we are still far from a democratic ideal of digital participation, be it institution-based or grassroots. The predominant contemporary vehicle for online democratic participation are social media platforms whose workings are inherently autocratic: they are closed-source and are controlled, in one way or another, by a central authority (the company owning the system) that does not respond to any democratic oversight from its users. Any form of digital democracy mediated by such platforms makes citizens inevitably and strongly dependent on whoever controls those systems \cite{bernholz2021digital,Ford:2021, MooreTambini:2018}.



\section{Closing words}

Digital democracy is increasingly perceived by citizens and public authorities as a technology that could offer a viable vehicle for more citizen-responsive policy making.
As such, it holds the promise of readdressing---at least in part---the growing societal discontent towards the way democracy is being practiced today. 
Yet, we know too little still about how social interaction and decision-making mediated by digital technology actually works: what are its pitfalls and how digital technology could be leveraged to foster democratic ideals in the decision-making processes of our societies.

\medskip

If digital democracy is to meet its promise, we need a much more robust understanding of it as a socio-technical system. In this paper we motivated and outlined a long-term research program of interdisciplinary science that can provide the missing foundations for the responsible development of a transparent, open and fit-for-purpose digital democratic infrastructure for our societies.


\appendix

\section*{Authors}

Coordinating authors (alphabetical order):
\begin{itemize}
    \item Davide Grossi (University of Groningen and University of Amsterdam, The Netherlands)
    \item Ulrike Hahn (Birkbeck College London, United Kingdom)
    \item Michael M\"as (Karlsruhe Institute of Technology, Germany)
    \item Andreas Nitsche (Interaktive Demokratie, Germany)
\end{itemize}

\noindent
Contributing authors (alphabetical order):
\begin{itemize}
    \item Jan Behrens (Interaktive Demokratie, Germany)
    \item Niclas Boehmer (Harvard University, USA)
    \item Markus Brill (University of Warwick, UK)
    \item Ulle Endriss (ILLC, University of Amsterdam, The Netherlands)
    \item Umberto Grandi (IRIT, Université Toulouse Capitole, France)
    \item Adrian Haret (MCMP, LMU Munich, Germany)
    \item Jobst Heitzig (Potsdam Institute for Climate Impact Research, Germany)
    \item Nicolien Janssens (EIPE, Erasmus University Rotterdam, The Netherlands)
    \item Catholijn M.\ Jonker (Technical University Delft \& University Leiden, The Netherlands)
    \item Marijn A.\ Keijzer (Institute for Advanced Study in Toulouse, France)
    \item Axel Kistner (Interaktive Demokratie, Germany)
    \item Martin Lackner (TU Wien, Vienna, Austria)
    \item Alexandra Lieben (University of California, Los Angeles, USA)
    \item Anna Mikhaylovskaya (University of Groningen, The Netherlands)
    \item Pradeep K.\ Murukannaiah (TU Delft, The Netherlands)
    \item Carlo Proietti (CNR, Italy)
    \item Manon Revel (Harvard University, USA)
    \item \'{E}lise Roum\'{e}as (University of Groningen, The Netherlands)
    \item Ehud Shapiro (Weizmann Institute of Science, Israel)
    \item Gogulapati Sreedurga (University of Edinburgh, United Kingdom)
    \item Bj\"orn Swierczek (Interaktive Demokratie, Germany)
    \item Nimrod Talmon (Ben-Gurion University, Israel)
    \item Paolo Turrini (University of Warwick, United Kingdom)
    \item Zoi Terzopoulou (GATE Lyon-Saint-Etienne, Jean-Monnet University, France)
    \item Frederik Van De Putte (Erasmus University Rotterdam, The Netherlands)
\end{itemize}

\section*{Acknowledgments}

The authors would like to acknowledge the generous support of the \href{https://nias.knaw.nl/}{Netherlands Institute for Advanced Studies in the Humanities and Social Sciences} (NIAS) and the \href{https://www.lorentzcenter.nl/}{Lorentz Center}.

Paolo Turrini acknowledges the support of the Leverhulme Trust for the Research Grant RPG-2023-050 titled ``Promoting Social Good Using Social Networks". Umberto Grandi acknowledges the support of the ANR JCJC project ``Social Choice on Social Networks" (ANR 18-CE23-0009-01).
Martin Lackner was supported by the Austrian Science Fund (FWF): P31890. Davide Grossi, Catholijn Jonker and Pradeep Murukannaiah acknowledge support by the \href{https://hybrid-intelligence-centre.nl}{Hybrid Intelligence Center}, a 10-year program funded by the Dutch
Ministry of Education, Culture and Science through the Netherlands
Organisation for Scientific Research (NWO).


\newcommand{\etalchar}[1]{$^{#1}$}

\end{document}